\documentclass[twocolumn]{aa}
\usepackage{graphicx}
\usepackage{txfonts}
\usepackage{graphics}
\usepackage{amssymb}                                                           
\usepackage{latexsym}
\begin{document}
   \title{Signatures of SN\,Ia in the galactic thick disk
\thanks{Based on
observations collected at the European Southern Observatory,
La Silla, Chile, Proposals \#65.L-0019(B) and 
67.B-0108(B)}}

   \subtitle{Observational evidence from
$\alpha$-elements in 67
 dwarf stars in the solar neighbourhood}

   \author{S. Feltzing, T. Bensby
          \and
          I. Lundstr\"om
          }

   \offprints{S. Feltzing, {\tt sofia@astro.lu.se}}

   \institute{Lund Observatory, Box 43, SE-221 00 Lund, Sweden\\
              \email{sofia,thomas,ingemar@astro.lu.se}
             }

   \date{Received ; accepted}

   \abstract{We present the first results of a larger study into the
stellar abundances and chemical trends in  long-lived dwarf stars in
the solar neighbourhood  that belong to (based on their kinematics)
the thin and thick galactic disk, respectively.  We confirm  that the
trends of $\alpha$-elements in the thin and thick disk are distinct
(this has previously been shown for Mg by Fuhrmann 1998, but e.g. Chen
et al. 2000 claimed the trends to follow smoothly upon each other).
We find that the thick disk show the typical
signature of contribution from SN\,Ia (i.e. the ``knee'') to the
enrichment of the interstellar gas out of which the later  generations
of thick disk stars formed.  The trend  starts out as [Mg/Fe]$\sim
0.35$ at [Fe/H]$\sim -0.7$ and continue on this level with increasing
[Fe/H] until $-0.4$ dex where a decline in [Mg/Fe] starts and steadily
continues down to 0 dex at solar metallicity. The same is true for the
other $\alpha$-elements (e.g. Si).  Using ages from the literature we
find that the  thick disk in the mean is older than the thin disk.
Combining our results with other observational facts we suggest that
the most likely formation scenario for the thick disk is, still, a
violent merger event. We also suggest that there might be tentative
evidence for diffusion of orbits in todays thin disk (based on
kinematics in combination with elemental abundances).
 } 
\authorrunning{Feltzing et al.}

   \maketitle
%

\section{Introduction}

Observational evidence revealed in the 1980's that the Galaxy has two
disk-like components: the thin and the thick disks (Gilmore \& Reid
1983). It is now also established that some, but not all, disk
galaxies possess a thin and a thick disk and that the presence of a
thick disk is often associated with mergers or interacting systems
(Schwarzkopf \& Dettmar 2000).

The stars in the thick disk have warm kinematics, ($\sigma_{\rm U},
\sigma_{\rm V},\sigma_{\rm W}$)=(67,38,35)  km s$^{-1}$, and appear to
all be old, e.g.  Fuhrmann (1998).  Its scale-height is 1000--1300 pc
which is  comparable to what is observed in other galaxies,
Schwarzkopf \& Dettmar (2000).  The thin disk is more confined to the
galactic  plane with a scale-height of 300 pc and the stars  have a
cooler kinematics,  ($\sigma_{\rm U},
\sigma_{\rm V},\sigma_{\rm W}$)=(35,20,16) km s$^{-1}$.  
Their metallicities are
in the mean higher than those in the thick disk and extend up to
super-solar values, Wyse \& Gilmore (1995).

The chemical evolution of the thick disk,  as traced by the stellar
abundances in long-lived dwarf stars in the solar neighbourhood, has
recently become a most active area of research, e.g. Fuhrmann (1998),
Prochaska et al. (2000),  Mashonkina \& Gehren (2001), Chen et al.
(2000), Gratton et al. (2000), and Tautvai\v{s}ien\.{e} et
al. (2001). However, the  conclusions reached by these studies point
in conflicting directions. While Gratton et al. (2000), Fuhrmann
(1998), and Mashonkina \& Gehren (2001) find that the star formation
in the thick disk lasted less than 1 Gyr, Prochaska et al. (2000)
infer a substantially longer time. Furthermore, three of the studies,
Fuhrmann (1998), Chen et al. (2000), and  Mashonkina \& Gehren (2001),
analyse the thin and thick disk  stars in a differential manner. Chen
et al. (2000) find that the chemical trends for the $\alpha$-elements
of the thin and thick disk stars, respectively, follow smoothly upon
each other, while Fuhrmann (1998) and Prochaska  et al. (2000) find
the trends for the thin and the thick disk to be clearly separate. All
the studies agree that the thick disk  is old.

Here we report our first findings of a larger study that we have
initiated in order to further characterize the two disks and also to
find out to what metallicities the thick disk extends. Former studies
have implied that the thick disk metallicity does not extend
significantly above --0.5 dex.  We discuss the implications of our
first results from our southern sample.  A northern sample is in
preparation.

   \begin{figure}
\resizebox{6cm}{!}
   {\includegraphics{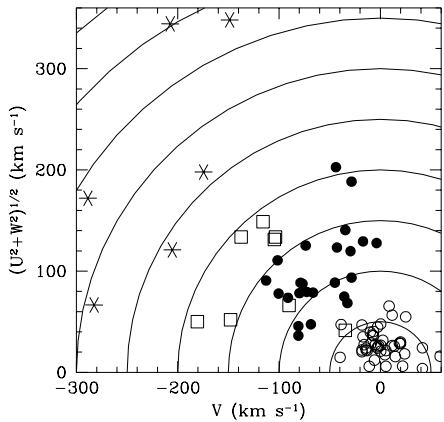}}
      \caption{Toomre diagram for our and Fuhrmann (1998) stars.
Our thin disk stars are denoted by $\circ$ and thick disk stars
with $\bullet$. Fuhrmann's thick disk stars are shown as
$\Box$ and his halo stars as $\ast$.
              }
         \label{kin.fig}
   \end{figure}

\section{Our data and results}

Stars were selected based on their kinematic properties.  In
particular we used the catalogue of more than 4000 stars compiled by
Feltzing \& Holmberg (2000) for their study of the moving group HR
1614.  We used those velocities in combination with the velocity
dispersions for the thin and thick disks, and the halo to calculate
probabilities that each star belong to either component.  We selected
only stars that had a very high probability to belong to either thin
or the  thick disk, respectively. 
When selecting stars we tried to cover the metal-poor
thin disk and  the thick disk with even numbers of stars per
metallicity bin (compare the sample selection in Edvardsson et
al. 1993). We also explicitly looked for thick disk stars with high
metallicities.  In this way we created a sample of 24  thick disk and
43 thin disk stars. The kinematic data are shown in
Fig. \ref{kin.fig}.  All stars in our sample have parallaxes with
errors $<$ 5 \% from Hipparcos and metallicity and temperature
estimates based on Str\"omgren  photometry, Feltzing et al. (2001).

The full details of our analysis as well as abundances of iron-peak
and other elements  will be presented in Bensby, et al.
(2002, in prep.).  Briefly, we perform a standard LTE analysis based
on equivalent widths, compare e.g. Edvardsson et al. (1993).  Model
atmospheres are created using the Uppsala MARCS code, Gustafsson et
al. (1974) and Edvardsson et al. (1993).

   \begin{figure}
\resizebox{\hsize}{!}
   {\includegraphics{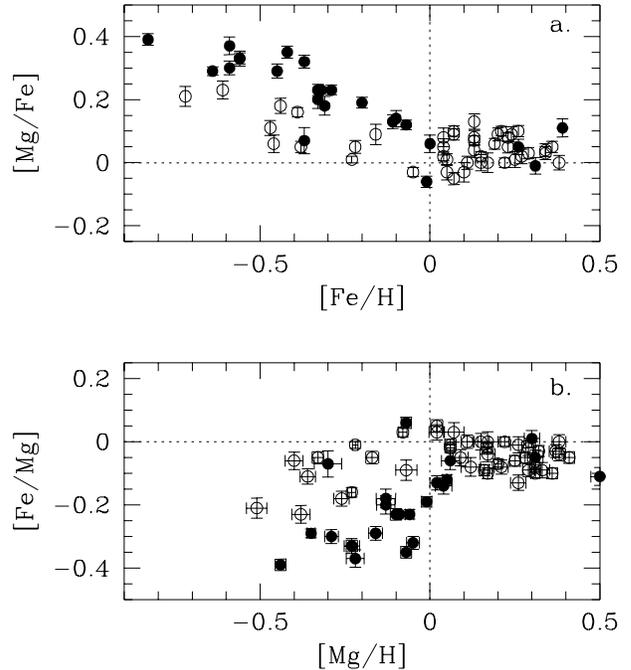}}
      \caption{Magnesium abundances for our program stars, error-bars
give the formal error in the mean ($=$ (line-to-line scatter)
$/\sqrt{N_{\rm lines}}$). The thick disk stars are shown as $\bullet$ and
the thin disk stars as $\circ$. Errors in [Fe/H] are smaller than the 
symbols.
              }
         \label{mg.fig}
   \end{figure}

\begin{figure}
\resizebox{\hsize}{!}
   {\includegraphics{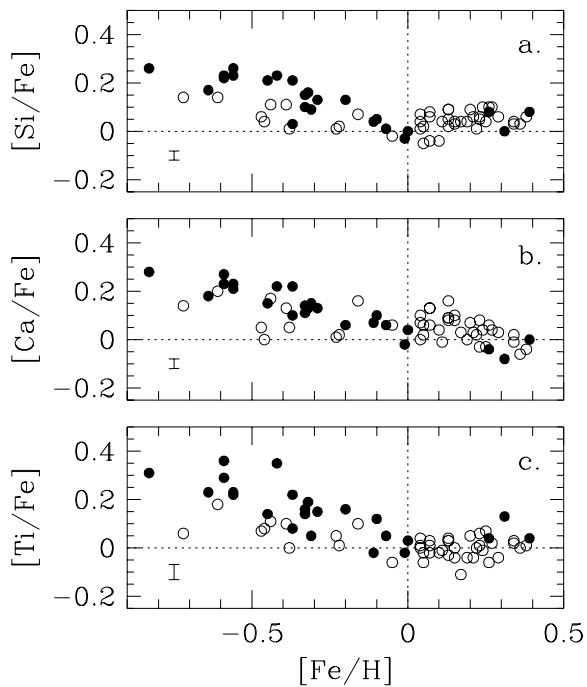}}
     \caption{Si, Ca, and Ti abundances for our thin disk stars,
$\circ$, and thick disk stars, $\bullet$.
Typical error-bars
are indicated at the bottom left and show the formal error in the mean
($=$ (line-to-line scatter)
$/\sqrt{N_{\rm lines}}$). 
              }
         \label{ab.fig}
   \end{figure}

Figures \ref{mg.fig} and \ref{ab.fig} show our abundance data. In Fig.
\ref{mg.fig} we also show the errors to let the reader
appreciate the {\sl internal} quality of the data.  Our two most
important findings are:

\begin{itemize}
\item Stars that belong to distinct kinematic populations show 
discrete trends as traced by all the  $\alpha$-elements.
\item The 
trends  for the thick disk stars show a clear signature of 
SN\,Ia contribution to the chemical enrichment (the ``knee'').
\end{itemize}

\noindent
Other findings are more model dependent (i.e. age determinations) or
are based on a small number of stars (2--5) and will require further
investigations. The most interesting of these are:

\begin{itemize}
\item  We find that in our sample of stars with well determined  ages
(see Feltzing et al. 2001) and with typical thick disk kinematics
have a mean age of  $12.1\pm 2.0$ Gyr. For the thin disk
stars  we get a mean age of $6.1 \pm 3.8$ Gyr.  Hence,
the two populations  do not overlap on the 1 $\sigma$
level.
\item Our sample contains two thin disk stars that appear very old.
\item Five stars with typical thick disk kinematics show typical thin
disk $\alpha$-abundances. 
\end{itemize}

\section{How did the galactic thick disk form?}
\label{sect:interpret}

%
%
%
%
%

\subsection{Possible scenarios}

How can thick disks form? Several  scenarios have been suggested and
were summarized in e.g. Gilmore et al. (1989).

\begin{itemize}

\item A slow, pressure supported collapse.

\item A fast collapse due to increased dissipation (e.g. enhanced
cooling by metals).

\item Violent heating by a 
merger, which puffs up an originally thin disk to the 
velocity dispersions of today's thick disk.

\item Accretion of thick disk material directly, e.g.
 satellites on suitable orbits.

\item Enhanced kinematic diffusion of the stellar orbits in the thin disk
eventually  producing a thickening.

\end{itemize}

\noindent
These various scenarios give definite predictions concerning the
metallicity, ages, and elemental abundances of the stars.  Some of the
scenarios have also been modelled in detail: the fast collapse was
studied by Burkert et al. (1992) and  the merger scenario by
e.g. Quinn et al. (1993).

In the first scenario vertical abundance gradients will build up and
the two disks will have overlapping age distributions.  The rapid
collapse, on the other hand, is  fast enough (see Burkert et al. 1992
for details) that indeed there will  be no time to build up gradients
in the thick disk (star formation ceases after about 400 Myr in the
thick disk). Further, the kinematics of the two components will be
different as well as their mean metallicities.

If the thick disk is due to violent heating by a merger then the  ages
of the thick disk stars will all be greater than the ages of the thin
disk stars. The abundance trends should, most likely, be
different. Note that the models (e.g. Quinn et al. 1993) do say
nothing about the formation scenario for the thin disk that is the
predecessor  of the thick disk. Thus there are no constraints on the
chemical trends we can expect to see in the thick disk. If on the
other hand material is accreted slowly over time from many smaller
satellites, on suitable orbits, ages and abundances will show a 
mixed picture.  The final mixture will depend on exactly when and how
much new gas (or stars) are accreted.

Lastly, if the thick disk is due to diffusion of stellar orbits then
the thick and thin disk stars come from the  same stellar
population. Thus they will share abundance trends as well as  age
distributions (although one can expect the thick disk to be mainly
old).  

\subsection{Observational facts}

From our differential study of the chemical trends in the thick and
thin  disks we find that the chemical trends in the two components are
distinct, that a clear signature of SNe\,Ia is present in the trends
for the thick disk, that the thick disk is old, and that there appear
to exist stars with typical thick disk kinematics that nevertheless
show typical thin disk elemental abundances.

Gilmore et al. (1995) showed that there is no discernible  vertical
metallicity gradient when going from 1 to 1.5 kpc above the  galactic
plane. At this height above the plane the thick disk dominates.

From a study of 110 edge-on disk galaxies Schwarzkopf \& Dettmar
(2000)  concluded that thick disks are common in galaxies in mergers
or in  interacting environments.

\subsection{Interpretation and discussion }

Our first key result is that the two disks show distinct abundance
patterns. This should rule out any model that predict that the disks
should have continuous distributions, e.g. only diffusion of  orbits
or only direct accretion of thin disk material.

Our second key result is that star formation went on long enough that
SN\,Ia started to enrich the gas out of which sub-sequent
generations of  stars formed in what is today the thick disk. Given a
model for SN\,Ia enrichment this sets a lower limit to how long time
star formation went  on for in the thick disk.

Often a value of 1 Gyr is quoted as the time-scale for SN\,Ia to
contribute to the chemical enrichment, however, this number is model
dependent. The most important parameter, given a model for SN\,Ia,
that sets this time-scale is the star-formation {\sl rate}. This is
illustrated in Matteucci (2001) (Fig. 5.7, see also references
therein) where  they show how, in a bursting scenario, the SN\,Ia rate
peaks significantly earlier than in e.g. the typical solar
neighbourhood (modelled to fit the over-all trends in the stellar
abundances, as our results demonstrate
such a simplification is not any longer
fully valid).

In fact it might be just possible to squeeze it in to the 400 Myr
found for the rapid collapse model by Burkert et al. (1992). In the
merger  scenario the star formation in what later became the thick
disk could have had at least a few Gyr to proceed (there are currently
no  observations apart from the thick disk stars themselves to
constrain the star formation time in the original thin disk) and thus
SN\,Ia may have contributed to the enrichment of the gas out of which
the later generations of stars in today's thick disk formed. In
summary, the presence of SN\,Ia signatures is not a strong enough
constraint to distinguish between a rapid collapse or a  violent
merger.

The lack of a vertical abundance gradient is often quoted as
supporting  a merger scenario. However, if there are gradients present
in the disk prior to the merger they will remain after the merger
(Quinn et al. 1993).  Furthermore, if the thin disk is around 7 Gyr
old and the merger took  about 1--3 Gyr to complete (Quinn et
al. 1993) then there would be ample  time left over to let the
original thin disk form through a slow, dissipative
collapse. Moreover, Burkert et al. (1992) show that in the rapid
collapse model it is  possible to have no vertical gradient over a
fairly large height.

We think that the extra-galactic evidence is important -- that thick
disks are associated with mergers, and would therefore advocate a
merger origin for the kinematic signature of the majority of the thick
disk stars. Obviously, some of today's thin disk stars may have got
their orbits diffused and would today look like thick disk stars if we
look at their orbits but like thin disk stars if we consider their
elemental abundances. The number of such stars should, however, be
small (Silk \& Wyse 1993).


Our results mainly agree with previous studies (see list in  Sect. 1),
i.e. that the thin and thick disks are distinct and the thick disk is
all old. However, all but one of the previous studies have found no
signatures of the SN\,Ia in the thick disk abundances.  From this a
short time-scale for the formation of the thick disk has generally
been derived (based on the, incorrect see above, assumption that
SN\,Ia timescale is 1 Gyr).   We argue that the reason for that
previous studies of the thick disk abundances have not seen this
signature is because they have deliberately excluded  more metal-rich
stars.  The reason for this is given (e.g. Fuhrmann 1998) as the
``fact'' that the thick disk should not have stars much more
metal-rich than $-0.5$ dex.  We think such an {\sl ad hoc} assumption
is ill-founded and preferred to include stars of all
metallicities. Our selection enabled us to find the presence of the
SN\,Ia signature.


As we have performed a differential analysis of the thin and thick
disk abundances our main results are robust. However, especially the
time-scale for star formation in the thick disk is not determined by
our  abundances (remember that SN\,Ia can contribute on a time scale
as  short as 0.5 Gyr or as long as 1--2 Gyr, Matteucci 2001).
Mashonkina \& Gehren (2001) have shown how the ratio of $r$- and
$s$-process elemental abundances  can be used to further constrain the
star formation time. Especially the presence of $s-$processed elements
(e.g. Ba) directly tells that  AGB stars have had time to contribute
to the enrichment of the interstellar medium out of which subsequent
stellar populations were born. For the  Bernkopf et al. (2001) stars
Mashonkina \& Gehren (2001) find that the $s$-process has not have had
time to contribute and hence the lifetime of the AGB  stars is longer
than the star formation time-scale in the thick disk. Since their
sample only contained stars with metallicities up to 
roughly --0.35 dex  their result
gives a lower, and not an upper, limit to the star formation time in
the thick disk.   Therefore it is important to  analyse $s$- and
$r$-process elements of more metal-rich thick disk stars. We are
currently doing this.
They also
found that the [Eu/Fe] started to decline at the highest metallicities
in their thick disk sample. This could indeed be an indication that 
SN\,Ia started to contribute to the chemical enrichment. However,
such a conjecture is based on the assumption that Eu is mainly
produced in the same sites as the $\alpha$-elements. 
The evidence from e.g. Mg for the same stars 
(see Bernkopf et al. 2002 Fig. 2) does not show any down-turn at the 
same metallicities.


Finally, the most important test would be to obtain stellar abundances
of dwarf stars that belong to the thick disk and are far away (around
1 kpc)  from the galactic plane. In such a sample no kinematic bias
would be  present as at these heights the thick disk dominates over
the thin disk  and stars could be chosen without selections based on
probabilities based on kinematics. Such observations, of e.g. the
Gilmore et al. (1995) stars, are now possible with the UVES
spectrograph on the VLT.

\section{Summary}

We find that the gas out of which today's thick disk stars have  formed
had been enriched by SNe\,Ia. This has implications for the  star
formation history of the thick disk. However, taking the  possibility
of a bursting star formation in the thick disk into  account it is not
impossible that such an enrichment could be  achieved in less than one
billion years (see e.g. Matteucci 2001).  This makes our  results
compatible with both a merger scenario (Quinn et al. 1993) and
(marginally) a fast dissipative collapse model (see Burkert et
al. 1992).

Further, we confirm previous results (e.g. Fuhrmann 1998, Prochaska et
al. 2000) that the thick disk is older than the thin and that the
abundance trends are discrete.

Based on our finding that there exist stars with typical thick disk
kinematics but with typical thin disk abundances and the fact that
extra-galactic studies find  thick  disks to be associated with
mergers the following, plausible, scenario for creating our observed
abundance patterns is suggested: today's thick disk originate from a
thin disk (which most likely was formed by rapid collapse,  e.g. no
gradients) that was later puffed up by a violent merger.  
After the gas had then settled into a thin disk the  stars
in today's thin disk were born. Some of the oldest of these stars have
later experienced significant kinematic diffusion and today have
kinematics typical of the thick disk but abundances typical of the
thin disk. That we found  an uncharacteristically large number of such
stars is because we  actively included stars with high metallicities
and thick disk kinematics.

\begin{acknowledgements}
TB acknowledges a travel grant from Fysiografiska s\"allskapet i Lund
for the first observing run at La Silla, Chile. SF acknowledges
computer grants from the same society.
      
\end{acknowledgements}


\begin{thebibliography}{}
\bibitem[]{} Bernkopf, J., Fiedler, A., \& Fuhrmann, K., 2002,
in Astrophysical Ages and Times Scales, ASP Conference Series Vol. 245. 
eds. by T. von Hippel, C. Simpson, and N. Manset., p. 207
\bibitem[]{} Burkert, A., Truran, J.W., \& Hensler, G., 1992, ApJ, 391, 651
\bibitem[]{} Chen, Y.Q., Nissen, P.E., Zhao, G., Zhang, H.W.,  \& Benoni, T., 
2000, A\&AS, 141, 491
\bibitem[]{} Edvardsson, B., Andersen, J., Gustafsson, B., Lambert, D.L.,
Nissen, P.E., \& Tomkin, J., 1993, A\&A 275, 101
\bibitem[]{} Feltzing, S., \& Holmberg, J., 2000, A\&A, 357, 153
\bibitem[]{} Feltzing, S., Holmberg, J., \& Hurley, J.R., 2001, A\&A, 377, 911
\bibitem[]{} Fuhrmann, K., 1998, A\&A, 338, 161
\bibitem[]{} Gilmore, G., Reid, N., 1983, MNRAS, 202, 1025
\bibitem[]{} Gilmore, G., Wyse, R.F.G., \& Jones, J.B., 1995, AJ, 109, 3
\bibitem[]{} Gilmore, G., Wyse, R.F.G., \& Kuijken, K., 1989, ARA\&A, 27, 555
\bibitem[]{} Gratton, R.G., Carretta, E., Matteucci, F., \& Sneden, S.,
2000, A\&A, 358, 671
\bibitem[]{} Gustafsson, B., Bell, R.A., Eriksson, K., \& Nordlund,  
{\AA}., 1975, A\&A 42, 407
\bibitem[]{} Mashonkina, L., \& Gehren, T.,  2001, A\&A, 376, 232
\bibitem[]{} Matteucci, F., 2001, The Chemical Evolution of the Galaxy, Kluwer
\bibitem[]{} Prochaska, J.X., Naumov, S.O., Carney, B.W., McWilliam, A., \& 
Wolfe, A.M., 2000, A\&A, 120, 2513
\bibitem[]{} Quinn, P.J., Hernquist, L., \& Fullagar, D.P., 1993, ApJ, 403, 74
\bibitem[]{} Schwarzkopf, U., \& Dettmar, R.-J., 2000, A\&A, 361, 451
\bibitem[]{} Silk, J., \& Wyse, R.F.G., 1993, Phys. Rep., 231, 293
\bibitem[]{} Tautvai\v{s}ien\.{e}, G., Edvardsson, B.,
 Tuominen, I., \& Ilyin, I., 2001, A\&A, 380, 578
\bibitem[]{} Wyse, R.F.G., \& Gilmore, G., 1995, AJ, 110, 2771
\end{thebibliography}
\end{document}